\documentclass[aps,prl,amsmath,amssymb,floatfix,twocolumn,amsmath,superscriptaddress,twocolumn,nofootinbib,tighten,letterpaper]{revtex4}
\usepackage{multirow}
\usepackage{subfigure}
\usepackage{color}
\usepackage{mathrsfs}
\usepackage{hyperref}
\usepackage[normalem]{ulem}
\usepackage{bm}

\usepackage{amssymb}   
\usepackage{amsmath}
\renewcommand\vec[1]{\ensuremath\boldsymbol{#1}} 

\usepackage{amsfonts, relsize, color}
\usepackage{graphics}
\usepackage{graphicx}
\usepackage{subfigure}
\usepackage{hyperref}
\usepackage{color}
\usepackage{comment}

\begin{document}

\title{Dispersive nodal fermions along grain boundaries in Floquet topological crystals}

\author{Daniel J. Salib}
\affiliation{Department of Physics, Lehigh University, Bethlehem, Pennsylvania, 18015, USA}

\author{Bitan Roy}~\thanks{Corresponding author:bitan.roy@lehigh.edu}
\affiliation{Department of Physics, Lehigh University, Bethlehem, Pennsylvania, 18015, USA}

\date{\today}

\begin{abstract}
Driven quantum materials often feature emergent topology, otherwise absent in static crystals. Dynamic bulk-boundary correspondence, encoded by nondissipative gapless modes residing near the Floquet zone center and/or boundaries, is its most prominent example. Here we show that topologically robust gapless dispersive modes appear along the grain boundaries, embedded in the interior of Floquet topological crystals, when the Floquet-Bloch band inversion occurring at a finite momentum (${\bf K}^{\rm Flq}_{\rm inv}$) and the Burgers vector (${\bf b}$) of the constituting array of dislocations satisfy ${\bf K}^{\rm Flq}_{\rm inv} \cdot {\bf b}=\pi$ (modulo $2 \pi$). Such nondissipative gapless states can be found near the center and/or edges of the Floquet Brillouin zone, irrespective of the drive protocol. We showcase these general outcomes for two-dimensional driven time-reversal symmetry breaking insulators. Promising experimental platforms hosting such dynamic topological dispersive bands in real materials are discussed.    
\end{abstract}

\maketitle

\emph{Introduction}.~Defects are ubiquitous in crystals, such as dislocations and grain boundaries. They are responsible for crystal melting that takes place through proliferation of lattice defects, which can be either pointlike, such as edge dislocations in two-dimensional (2D) crystals or extended, such as screw dislocations and grain boundaries. Furthermore, line defects can often be constructed by stacking point defects. For example, an array of edge dislocations gives rise to a grain boundary~\cite{book1}. In recent time, such geometric lattice defects have gained a revived interest in the context of topological quantum materials~\cite{Hasan-Kane-RMP, Qi-Zhang-RMP}. Under conducive environments, they can harbor robust topological modes in their vicinity that are, most importantly, immune to interface contamination~\cite{ran-zhang-vishwanath, teo-kane, nagaosa:dislocation, juricic:defectPRL, hughesyaoqi, slagerjuricic:dislocationPRB2014, youchohughes, juricic:grainboundary, hamasi:dislocation, nayak:dislocation, jian:dislocation, queiroz:dislocation, nagroy:floquetdislocation, roy-juricic-dislocation, panigrahi:NHdislocation, panigrahi:commphysdislocation, veluryhughes:linedefect, sanjibroy:dynamicdislocation, sanjibroy:dislocationSC, amundsen-juricic}. As lattice defects locally break the translational symmetry in the bulk of crystals, topological phases harboring such defect modes are thus named translationally active. While lattice defects in static topological materials have been scrutinized thoroughly over the span of last few years, their role as smoking gun probe of dynamic topological phases is still at its infancy~\cite{nagroy:floquetdislocation, sanjibroy:dynamicdislocation}.

\begin{figure}[t!]
\includegraphics[width=1.00\linewidth]{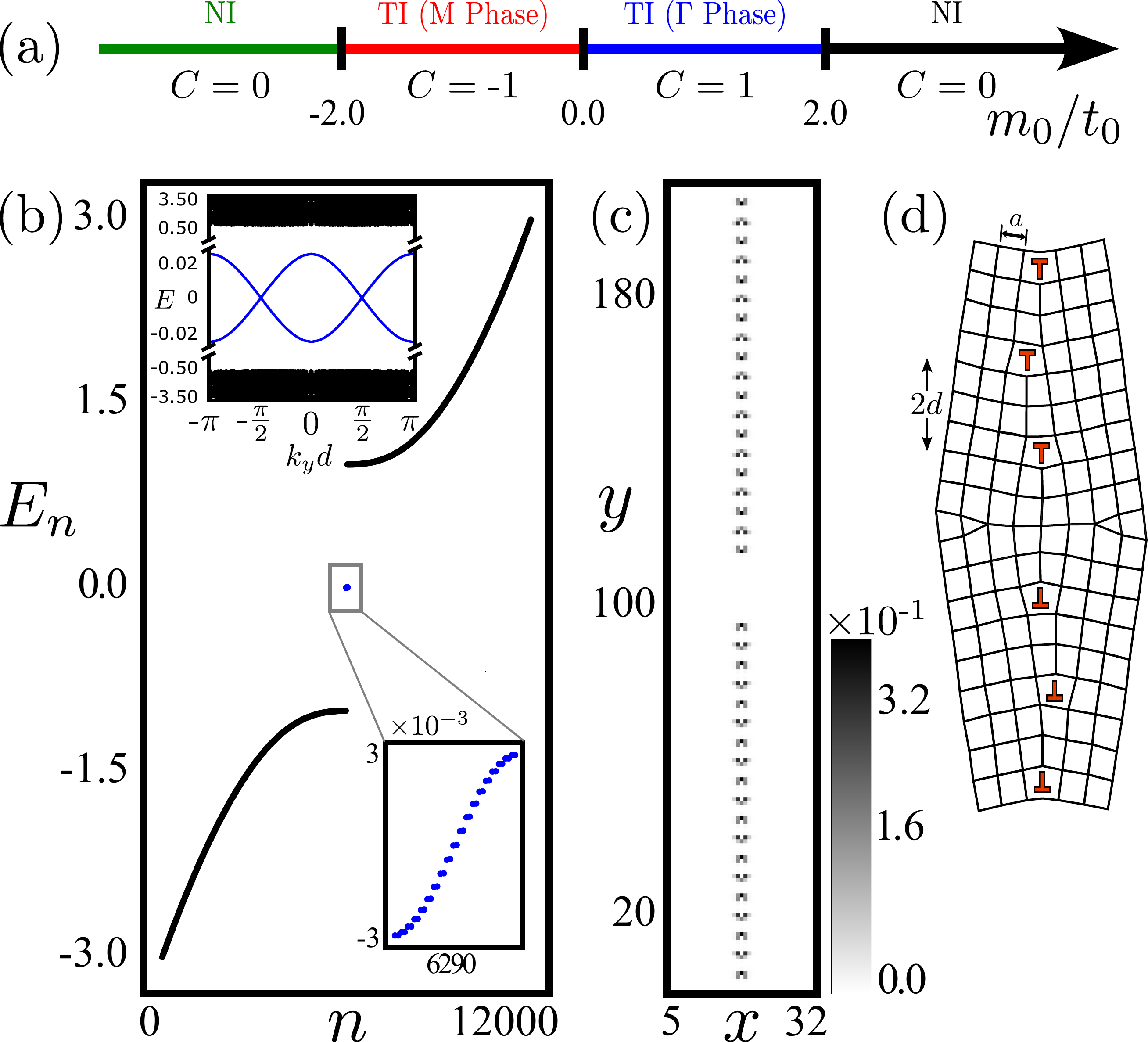}
\caption{(a) Phase diagram of the static Hamiltonian $H_{\rm stat}$ [Eq.~(\ref{eq:dvector})] in terms of the Chern number (C) [Eq.~(\ref{eq:chernnumber})]. (b) Energy ($E_n$) spectra of $H_{\rm stat}$ in a periodic system with a pair of grain-antigrain boundary each containing 10 (anti)dislocations, with the Burgers vector ${\bf b}=\pm a \hat{\bf e}_1$ for $t_1=t_0=1$ and $m_0=-1.5$, such that the system is in the translationally active ${\rm M}$ phase. Here $n$ is the energy eigenvalue index, and the system contains total 6290 sites. One-dimensional dispersive states (blue dots in lower inset) then appear along the line defects, as can be seen from their Fourier transformation as a function of the conserved momentum $k_y$ (upper inset). (c) Local density of states of these modes is highly localized along the line defects. (d) An illustration of a single grain-antigrain boundary pair with three (anti)dislocations. Their cores are shown in red. Throughout, the distance between two successive dislocation cores is $2d$.        
}~\label{fig:static}
\end{figure}

Here we showcase emergence of one-dimensional (1D) dynamic topological dispersive nodal fermions along the grain boundary of a 2D Floquet topological crystal. For simplicity, we consider a system that in the static limit features both topological insulators (TIs) and atomic or normal insulators (NIs) at the cost of the time-reversal symmetry. In particular, when the band inversion of the TI takes place at a momentum ${\bf K}_{\rm inv}$ in the Brillouin zone such that along with the Burgers vector of the underlying constituting dislocation ${\bf b}$, it satisfies ${\bf K}_{\rm inv} \cdot {\bf b}=\pi$ (modulo $2 \pi$)~\cite{ran-zhang-vishwanath}, 1D dispersive states appear along the grain boundary~\cite{juricic:grainboundary}. They form a miniband within the bulk topological band gap along the line defect. See Fig.~\ref{fig:static} and Fig.~S1 of Supplemental Information (SI). It is worth noting that this topological criterion also applies to static higher-order topological crystals, where dispersive bands typically emerge at finite energies while remaining protected by symmetries~\cite{danHOTGB}.

\begin{figure*}[t!]
\includegraphics[width=1.00\linewidth]{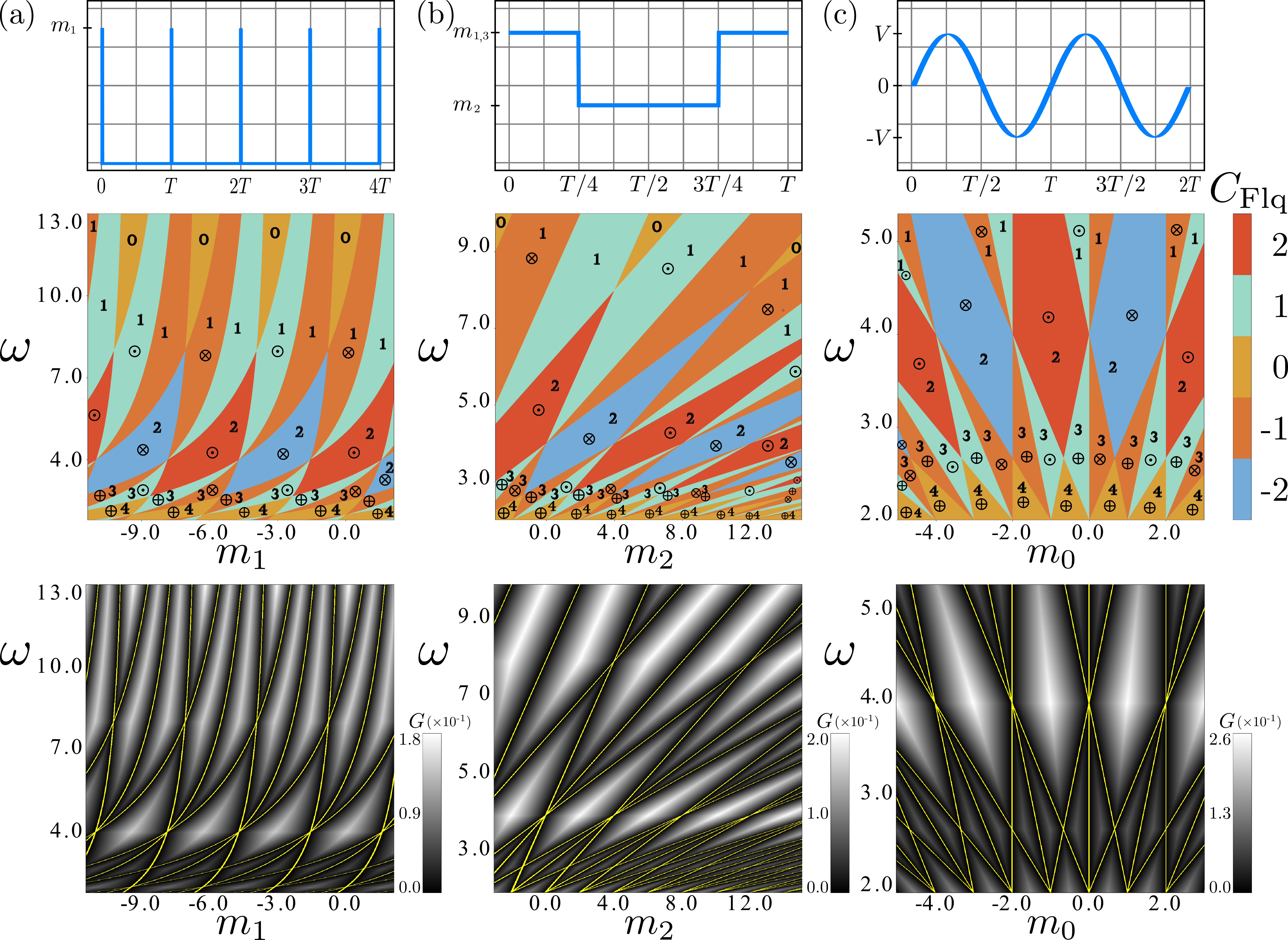}
\caption{Global phase diagrams of time reversal odd insulators, subject to (a) kick [Eq.~(\ref{eq:kickdrive})], (b) step [Eq.~(\ref{eq:stepdrive})] and (c) sinusoidal [Eq.~(\ref{eq:sinusoidal})] drives, schematically shown in the corresponding upper panel, with $T$ as the period of the drives. Here $\omega=2\pi/T$ is the drive frequency. Phases are colored (numbered) according to the Floquet Chern number $C_{\rm Flq}$ (total number of edge modes in the Floquet Brillouin zone or winding number ${\mathcal W}$~\cite{Flq:Thr3}). Insulators supporting normal, anomalous and mixed (with both normal and anomalous) one-dimensional dynamic gapless fermionic modes along the grain boundary are marked by $\odot$, $\otimes$ and $\oplus$, respectively. See Figs.~\ref{fig:kick}-\ref{fig:sinusoidal}, and Supplemental Information. Here we set $t_1=t_0=1$, and (a) $m_0=3$, (b) $m_0=3$, $m_1=m_3=2$, and (c) $V=3$. Lower panel: A heat-map of the bulk gap ($G$) in the phase diagram shown in the middle panel, showing that the bulk gap vanishes only at the phase boundaries (yellow dashed line) between topologically distinct insulators.      
}~\label{fig:phase}
\end{figure*}

\begin{figure}[t!]
\includegraphics[width=1.00\linewidth]{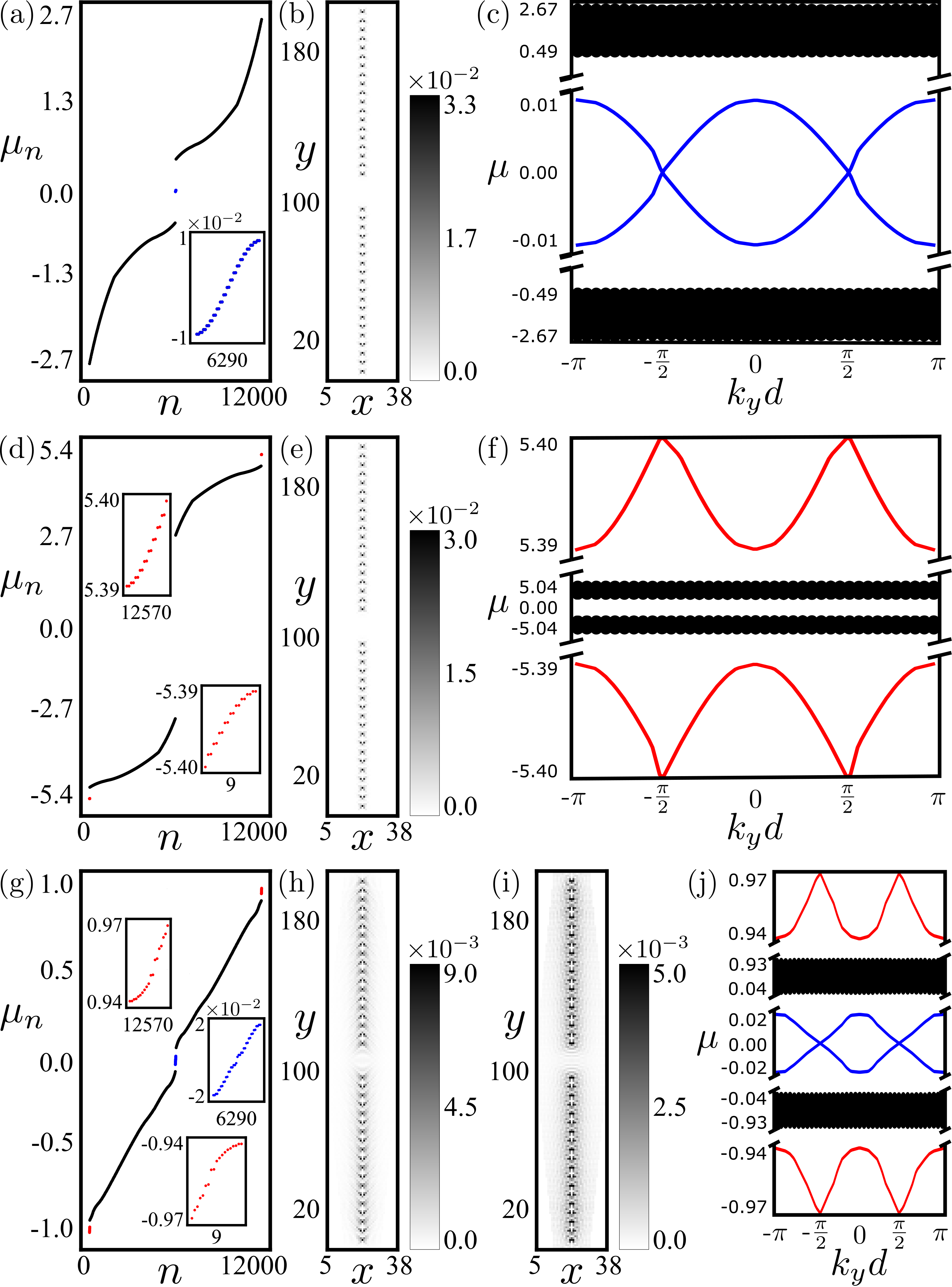}
\caption{Dynamic dispersive modes along the grain-antigrain boundary stemming from a kick drive [Eq.~(\ref{eq:kickdrive})]. (a) Quasienergy ($\mu_n$) spectra in a system with periodic boundary conditions and a grain-antigrain boundary pair, each containing 10 (anti)dislocations, for $\omega=12.8$, $m_1=-1.8$ producing normal dynamic nodal fermions near the Floquet zone center (blue dots). (b) Their local density of states (LDOS) is highly localized along the line defects. (c) Fourier transformation of the same set of states confirms their dispersive nature. Panels (d), (e) and (f) are similar to (a), (b) and (c), respectively, however for $\omega=10.8$, $m_1=-5.2$ giving rise to anomalous dynamic nodal fermions (red dots) near the Floquet zone boundary. Panel (g) is similar to (a), but for $\omega=1.94$, $m_1=-4.5$ hosting both normal (blue) and anomalous (red) gapless fermionic modes, for which the LDOS are respectively shown in (h) and (i). (j) Fourier transformation pins their dispersive nature near the Floquet zone center and boundary. We set $m_0=3$ and $t_1=t_0=1$. In panels (a), (d) and (g), $n$ is the quasienergy eigenvalue index, and the system contains total 6290 sites.      
}~\label{fig:kick}
\end{figure}

\begin{figure}[t!]
\includegraphics[width=1.00\linewidth]{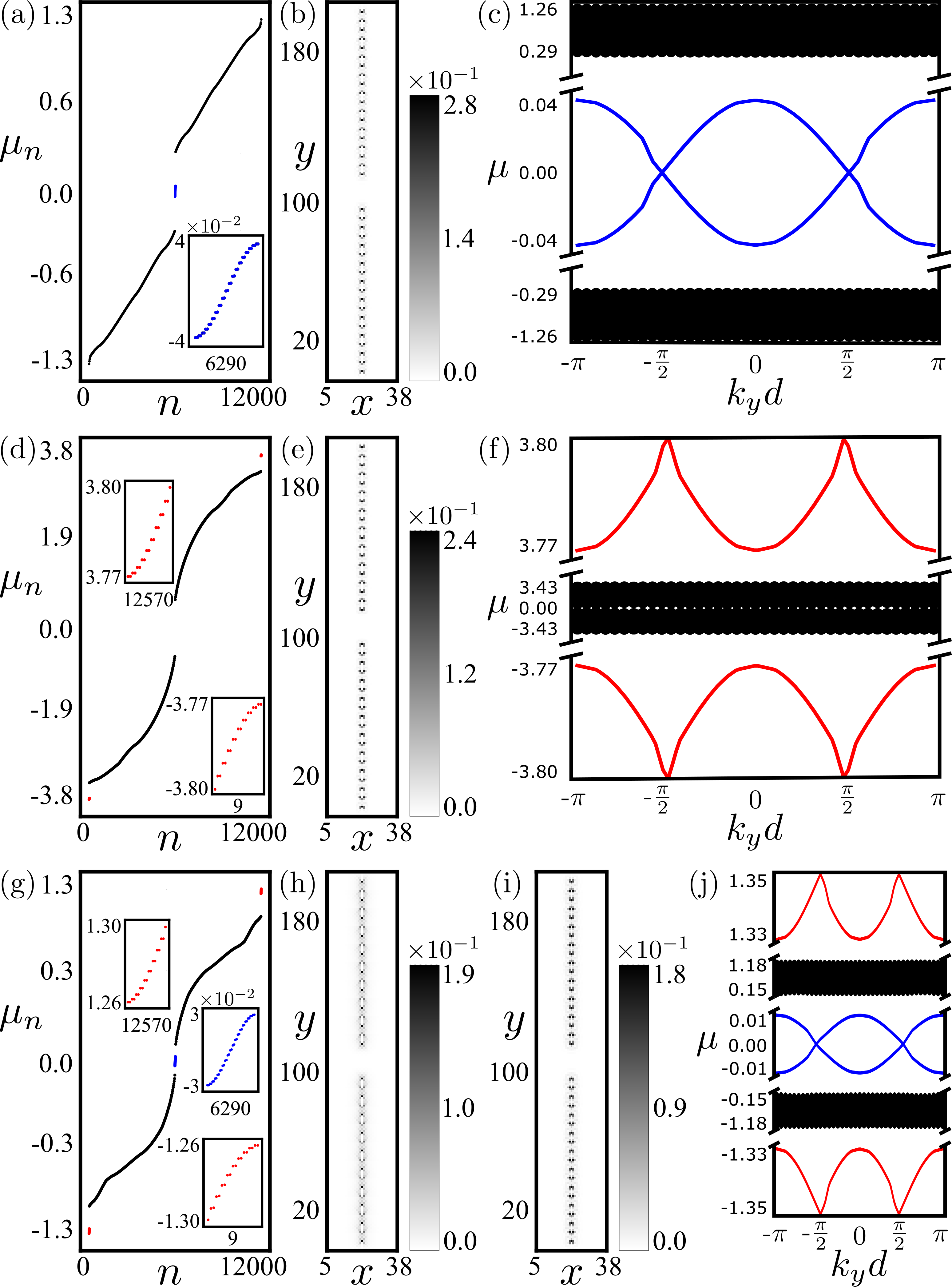}
\caption{Dynamic dispersive modes along the grain-antigrain boundary with a step drive [Eq.~(\ref{eq:stepdrive})]. (a) Quasienergy ($\mu_n$) spectra in a system with a grain-antigrain boundary pair, each containing 10 (anti)dislocations, and periodic boundary conditions for $\omega=3.5$, $m_2=-2.5$ producing normal dynamic nodal fermions near the Floquet zone center (blue dots). (b) Their local density of states (LDOS) displays strong localization along the line defects. (c) Fourier transformation of the same set of states, confirming their dispersive nature. Panels (d), (e) and (f) are similar to (a), (b) and (c), respectively, however for $\omega=7.6$, $m_2=-2.8$ supporting anomalous dynamic dispersive modes (in red) near the Floquet zone boundary. (g) is similar to (a), but for $\omega=2.7$, $m_2=-2.9$ hosting both normal (blue) and anomalous (red) dispersive fermionic modes, whose LDOS are respectively shown in (h) and (i). (j) Their Fourier transformation shows dispersive nature near both the Floquet zone center and boundary. Here we set $m_0=3$, $m_1=m_3=2$ and $t_1=t_0=1$. In panels (a), (d) and (g), $n$ is the quasienergy eigenvalue index, and the system contains total 6290 sites.
}~\label{fig:step}
\end{figure}

\begin{figure*}[t!]
\includegraphics[width=1.00\linewidth]{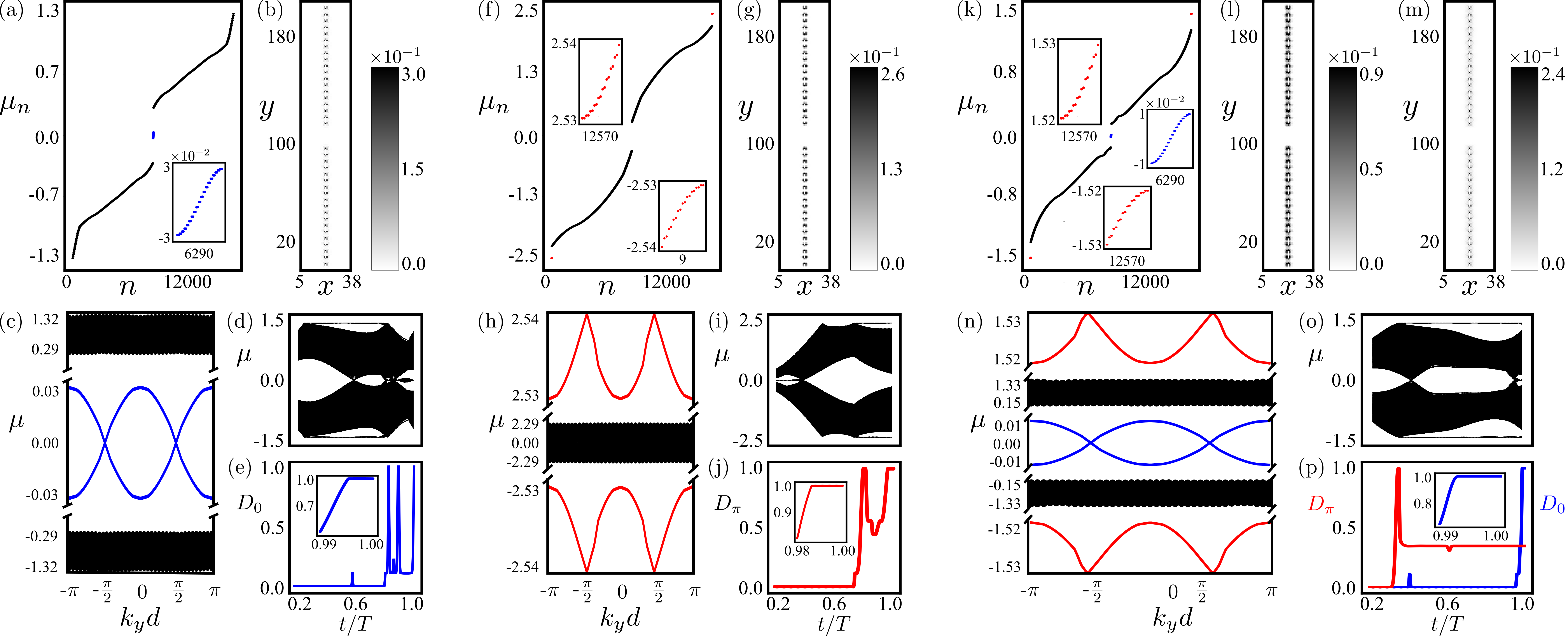}
\caption{Dynamic dispersive band along the grain-antigrain boundary with a sinusoidal drive [Eq.~(\ref{eq:sinusoidal})]. (a) Quasienergy ($\mu_n$) spectra in a system with periodic boundary conditions and a grain-antigrain boundary pair, each containing 10 (anti)dislocations, for $\omega=3.10$, $m_0=2.42$, yielding a normal gapless band near the Floquet zone center (blue dots). Their (b) local density of states (LDOS) is highly localized along the line defects and (c) Fourier transform depicts the dispersive nature. Time evolution of (d) quasienergy spectra and (e) normalized density of states (NDOS) near Floquet zone center ($D_0$) in a system with 5 (anti)dislocations forming the grain-antigrain boundary pair for the same parameters as in (a). Panels (f)-(j) are similar to (a)-(e), respectively, but for $\omega=5.08$, $m_0=-2.79$, producing an anomalous (red) dispersive band along the line defects. Here $D_\pi$ denotes the NDOS near the Floquet zone boundary. Panel (k) is similar to (a), but for $\omega=3.06$, $m_0=1.33$, giving rise to both normal (blue) and anomalous (red) dispersive nodal fermions along the line defects. Their LDOS are respectively shown in (l) and (m), while (n) Fourier transformation shows dispersive nature near both the Floquet zone center and boundary. Time evolution of (o) quasienergy spectra and (p) NDOS near the Floquet zone center (blue) and boundary (red) in a system as in (d). We set $V=3$ and $t_1=t_0=1$. In panels (a), (f) and (k), $n$ is the quasienergy eigenvalue index, and the system contains total 6290 sites. 
}~\label{fig:sinusoidal}
\end{figure*}

Irrespective of the periodic drive protocol, we show that such a system features a plethora of topological and normal insulating phases in the dynamic realm even when its static counterpart describes a NI, imprinted within the corresponding global phase diagrams shown in Fig.~\ref{fig:phase}, depending on the drive frequency ($\omega$) and its amplitude. The time translational symmetry then gives birth to the Floquet Brillouin zone (FBZ) within the quasienergy $\mu \in \left(-\omega/2,\omega/2 \right)$~\cite{Flq:Thr1, Flq:Thr2, Flq:Thr3, Flq:Thr4, Flq:Thr5}. So, all the quasienergy spectra are shown within the range of $\left(-\omega/2,\omega/2 \right)$ with the $\omega$ values specified explicitly, whenever appropriate. The FBZ is distinct from the regular Brillouin zone in the space of spatial momenta ($\vec{k}$). The inversion of the Floquet-Bloch bands, taking place at spatial momentum ${\bf K}^{\rm Flq}_{\rm inv}$, thus can occur near the FBZ center and/or near its boundaries. Under this circumstance, the ${\bf K} \cdot {\bf b}$ rule for the dispersive grain boundary modes extends to dynamic systems, but in terms of ${\bf K}^{\rm Flq}_{\rm inv}$. And we display the appearance of nondissipative dispersive (a) normal 1D dynamic modes around the FBZ center and zero quasienergy, (b) anomalous 1D dynamic modes around the FBZ boundaries and quasienergies $\pm \omega/2$, and (c) mixed 1D dynamic modes, simultaneously featuring both normal and anomalous dynamic dispersive nodal fermions along the grain boundary. These generic outcomes are explicitly shown for the kick [Fig.~\ref{fig:kick}], step [Fig.~\ref{fig:step} and Fig.~S2 of the SI] and sinusoidal [Fig.~\ref{fig:sinusoidal}] drives. They are qualitatively similar for small angle grain boundaries (SAGBs) and their large angle counterparts. See Fig.~S3-S5 of the SI. It should be noted that the lattice with grain and anti-grain boundaries does not possess any crystal symmetry, such as reflections about $x$ and $y$ axes, inversion, or four-fold rotation about $z$ axis. See Fig.~\ref{fig:static}(d). Thus, the existence of the dispersive modes along the line defects rests on a single and robust topological criterion ${\bf K}^{\rm Flq}_{\rm inv} \cdot {\bf b}=\pi$ (modulo $2\pi$). In the presence of isolated dislocation lattice defects, all results remain valid; however, the dispersive bands are replaced by localized modes around such pointlike crystal defects~\cite{nagroy:floquetdislocation}.

In noninterating and isolated Floquet crystals (decoupled from any dissipative bath), all the quasimodes (including the gapless dispersive ones near the grain boundaries) are nondissipative, due to their conserved quasienergies, as there is no gain or loss of energy in the system~\cite{Flq:Thr1, Flq:Thr2, Flq:Thr3, Flq:Thr4, Flq:Thr5}. Inter-particle interactions cause dissipation leading to heating or thermalization, which however occurs beyond a time scale $\tau^\star \sim \exp[\omega/\text{interaction strength}]$~\cite{thermalization:Th1, thermalization:Th2, thermalization:Th3}. Therefore, in the high frequency regime and/or weakly interacting systems, the thermalization takes place only at sufficiently long time, and at any time scale shorter than $\tau^\star$, the system is well approximated by the effective Floquet Hamiltonian and its nondissipative states, now describing the transient dynamics in a pre-thermalized state. Therefore, our ``dissipationless" grain boundary modes can be observed in real systems as transient states at time scale shorter than $\tau^\star$. Recently, Floquet-Bloch states has been observed as transient states in graphene (a weakly interacting system) within an experimentally achievable time scale that is shorter than $\tau^\star$~\cite{thermalization:Ex1, thermalization:Ex2}. However, these couplings are absent in classical dynamic metamaterials, featuring bosonic or classical analogues of topologically protected dispersive grain boundary modes.

\emph{Model}.~A pedagogical overview on the static system and grain boundaries therein will benefit the forthcoming discussion on the role of such line defects in Floquet crystals. The Hamiltonian for the static system is given by $H_{\rm stat}={\boldsymbol \tau} \cdot {\bf d}(\vec{k})$~\cite{qiwuzhang}, where $\vec{k}$ is spatial momenta,
\allowdisplaybreaks[4] 
\begin{equation}~\label{eq:dvector}
    \mathbf{d}(\vec{k}) = \biggl(t_1 \sin(k_x a), t_1 \sin(k_y a), m_0 -  t_0 \sum_{j=x,y} \cos(k_j a) \biggr),
\end{equation}
and $a$ is the lattice spacing. Vector Pauli matrix ${\boldsymbol \tau}$ operates on orbitals. Two component spinor reads $\Psi^\top(\vec{k})=[c_{+}(\vec{k}),c_{-}(\vec{k})]$, where $c_\tau(\vec{k})$ is the fermionic annihilation operator on orbital with parity $\tau=\pm$ and momentum $\vec{k}$. This model supports TIs (NIs) within the parameter range $|m_0/t_0|<2$ ($|m_0/t_0|>2$). Within the topological regime, the inversion of the Bloch bands (${\bf K}_{\rm inv}$) takes place near the $\Gamma=(0,0)$ and the ${\rm M}=(1,1)\pi/a$ points of the Brillouin zone for $0<m_0/t_0<2$ and $-2<m_0/t_0<0$, respectively, and are named the $\Gamma$ phase and ${\rm M}$ phase. The band inversion momentum can be recognized from the band structure of $H_{\rm stat}$ in a semi-infinite system with only $k_x$ or $k_y$ as a good quantum number. The edge modes then cross the zero energy at momentum ${\bf K}_{\rm inv}$. See Sec.~S1 and Fig.~S1 of the SI. These two TIs are also characterized by distinct first Chern number ($C$) computed in the following way~\cite{TKKN, fukui}. We neglect the particle-hole asymmetry (proportional to two-dimensional identity matrix $\tau_0$), as it does not play any role in topology as long as the system is a bulk insulator.

We consider a discrete 2D Brillouin zone, containing reciprocal lattice points $\mathbf{k_\ell} = (k_{j_1}, k_{j_2})$, where $k_{j_\mu} = (2 \pi j_\mu/N_\mu) - \pi$, $j_{\mu} = 0, \ldots ,N_{\mu}-1$ and $\mu=1,2$. For simplicity, here we take $N_1=N_2=N$. The Brillouin zone is restricted within $\vert k_{j_\mu} \vert < \pi$. A $\textrm{U(1)}$ link variable for $\mathbf{k_\ell} \rightarrow \mathbf{k_\ell}+\bm{\hat\mu}$ is defined as $U_{\mu}(\mathbf{k_\ell}) \equiv \langle{n(\mathbf{k_\ell})}\vert{n}(\mathbf{k_\ell}+\bm{\hat\mu})\rangle/A_\mu(\mathbf{k_\ell})$, where $\vert{n}(\mathbf{k_\ell})\rangle$ is the normalized eigenstate of band $n$ of $H_{\rm stat}$ at momentum $\mathbf{k_\ell}$, $A_\mu(\mathbf{k_\ell})={\vert\langle{n(\mathbf{k_\ell})}\vert{n}(\mathbf{k_\ell}+\bm{\hat\mu})\rangle\vert}$ and $\bm{\hat\mu} = (2\pi/N)\bf{\hat{e}_{\mu}}$. A counter-clockwise path around a unit plaquette in the reciprocal space is then represented by
\allowdisplaybreaks[4] 
\begin{equation}
    P_{12}(\mathbf{k_\ell}) = U_1(\mathbf{k_\ell})U_2(\mathbf{k_\ell}+\mathbf{\hat{1}})U_1(\mathbf{k_\ell}+\mathbf{\hat{2}})^{-1}
    U_2(\mathbf{k_\ell})^{-1},
\end{equation}
yielding lattice field strength $F_{12}(\mathbf{k_\ell}) = \ln{P_{12}(\mathbf{k_\ell})}$ with $-\pi \leq -i F_{12}(\mathbf{k_\ell}) \leq \pi$. The corresponding Chern number 
\allowdisplaybreaks[4] 
\begin{equation}~\label{eq:chernnumber}
    C_n = \frac{1}{2\pi{i}}\sum_{\ell}F_{12}(\mathbf{k_\ell})
\end{equation}
typically converges for $N=30$. Throughout, we compute it for the valence band and set $C_n=C$ for brevity, leading to $C=+1$ ($-1$) in the $\Gamma$ (${\rm M}$) phase.

When an edge dislocation is introduced in an otherwise square lattice system through the Volterra cut-and-paste procedure, an electron encircling the core of such lattice defect picks up a hopping phase $\Phi_{\rm dis}={\bf K}_{\rm inv} \cdot {\bf b}$ (modulo $2 \pi$). The Burgers vector ${\bf b}$ measures the missing translation around the defect core across the line of missing atoms. Here we take ${\bf b}=a \hat{\bf e}_1$. Then, in the ${\rm M}$ phase $\Phi_{\rm dis}=\pi$, while $\Phi_{\rm dis}=0$ in the $\Gamma$ phase. Thus only in the ${\rm M}$ phase a  nontrivial $\pi$-flux threads the dislocation core and the system supports a localized topological mode in its close vicinity~\cite{juricic:defectPRL}, pinned at zero energy due to an antiunitary particle-hole symmetry of $H_{\rm stat}$, namely $\{H_{\rm stat}, \Theta \}=0$ where $\Theta =\tau_1 {\mathcal K}$ and ${\mathcal K}$ is the complex conjugation~\cite{roysolo:antiunitary}.

Once a grain boundary is created from the array of such edge dislocations, tunneling among the zero energy modes bound to each dislocation core causes hybridization among them. As a result, a 1D dispersive miniband develops within and separated from the bulk band gap~\cite{juricic:grainboundary}. Such 1D topological dispersive modes reside along the grain boundary. Their dispersive nature can be anchored from the corresponding Fourier transformation in terms of the conserved momentum $k_y$ along the grain boundary for ${\bf b}=a \hat{\bf e}_1$. When the dislocation modes hybridize, besides maintaining high localization at the defect core they also develop comparable spectral weight in between them. For this reason, while performing their Fourier transformation, we denote the distance between two successive defect cores by $2d$. These results are summarized in Fig.~\ref{fig:static}. The grain boundary is characterized by the angle $\theta=\sin^{-1}(a/d)$. For SAGBs we set $d=5a$, yielding $\theta=11.54^{\circ}<15^{\circ}$. Throughout, we consider a single grain-antigrain boundary pair to impose periodic boundary conditions in all directions. The ends of the grain and antigrain boundaries are separated by a distance $20 a$ for SAGBs, so the modes bound to them do no overlap. Appearance of the dispersive fermionic modes can be observed when (anti)grain boundary contains more than 5-7 (anti)dislocation cores, such that hybridization takes place among a large number of defect modes. The width of the system in the $x$ direction is kept sufficiently large such that the (anti)grain boundaries are buried in the deep interior of the crystal, although we impose periodic boundary conditions in all directions.

\emph{Driven system}.~With the stage being set, we now focus on the same system under periodic drives. In particular, here we consider three drives. (a) A kick drive with 
\allowdisplaybreaks[4] 
\begin{equation}~\label{eq:kickdrive}
V(t) = m_1 \tau_3 \sum_{n=-\infty}^{\infty}\delta(t - nT)
\end{equation}
where $n$ is an integer, (b) a step drive given by
\allowdisplaybreaks[4] 
\begin{equation}~\label{eq:stepdrive}
    V(t) = \begin{cases}
  m_1 \tau_3 & nT < t < nT + T/4, \\
  m_2 \tau_3 & nT + T/4 < t < nT + 3T/4 \\
  m_3 \tau_3 & nT + 3T/4 < t < (n+1)T,
\end{cases},
\end{equation}
where we set $m_1=m_3$ only for simplicity and taking $m_1 \neq m_3$ does not change the results qualitatively [see Fig.~S2 of the SI], and (c) a sinusoidal drive with 
\allowdisplaybreaks[4] 
\begin{equation}~\label{eq:sinusoidal}
    V(t) = V\cos(2 \pi t/T)\tau_3.
\end{equation}
The drive frequency is $\omega=2 \pi/T$. See Fig.~\ref{fig:phase}(top row). The corresponding time ordered (TO) Floquet unitary operator is 
\allowdisplaybreaks[4]  
\begin{equation}~\label{eq:unitarydefinition}
    U(\mathbf{k}, t) = \text{TO}\left(\exp\left[-i\int_{0}^{t}\left[ H_{\rm stat}(\mathbf{k}) + V(t)\right] dt\right] \right).
\end{equation}
For sinusoidal drive, $U(\mathbf{k}, t)$ is computed employing the Trotter-Sujuki approximation~\cite{trotter, sujuki}. We are mostly interested at the stroboscopic time $t=T$, at which the effective Floquet Hamiltonian is given by 
\allowdisplaybreaks[4] 
\begin{equation}
H_{\text{Flq}}(\mathbf{k}) = i\ln(U(\mathbf{k}, T))/T=\mathbf{d}_{\text{Flq}}(\mathbf{k},T)\cdot\bm{\tau}.  
\end{equation}
For the explicit form of $\mathbf{d}_{\text{Flq}}(\mathbf{k},T)$ see Sec.~S2 of the SI. The global phase diagram of such a system can be constructed in terms of the Chern number for $H_{\text{Flq}}(\mathbf{k})$, employing the method discussed earlier for $H_{\rm stat}$. See Fig.~\ref{fig:phase} (middle row). In the high-frequency regime, each phase occupies a larger parameter space in the drive amplitude and frequency plane, whereas in the low-frequency regime, the parameter space for individual phases shrinks due to increased system complexity. However, in contrast to the phase diagrams for the kick and step drives, the Chern number flips under $m_0 \to -m_0$ for the sinusoidal drive. This occurs because, when during the first half of the drive period, $V(t)$ is positive (negative), during the second half, it becomes negative (positive). Consequently, the Chern number reverses under $m_0 \to -m_0$, as this transformation causes $\mathbf{d}^{3}_{\text{Flq}} (\vec{k},T) \to -\mathbf{d}^{3}_{\text{Flq}} (\vec{k},T)$.

We also show that the bulk gap ($G$) in the Floquet system vanishes only at the phase boundaries between topologically distinct insulating phases in Fig.~\ref{fig:phase} (bottom row) in both high and low frequency regimes. Irrespective of the drive frequency and amplitude, as the bulk Floquet spectrum is always gapped and gapless modes (when exist) live only along the 1D grain boundary. Thus, the noninteracting and isolated Floquet system is always in the dissipationless regime~\cite{Flq:Thr2}.

When periodically driven, the inversion of the Floquet-Bloch bands can take place near the center and/or boundaries of the FBZ. Recall, a static TI with the band inversion at the $\Gamma$ ($\rm M$) point leads to $C=+1$ ($-1$), tunable by the mass term proportional to $\tau_z$. The same conclusions hold for the Floquet insulators when mass term becomes time periodic with the the band inversion occurring near the FBZ center. By contrast, the correspondence between the Chern number and band inversion momentum gets reversed when it takes place near the boundaries of the FBZ. The net Chern number for Floquet insulators is given by $C_{\rm Flq}=C_{\rm FZC} + C_{\rm FZB}$, where $C_{\rm FZC}$ ($C_{\rm FZB}$) denotes the Chern number stemming from the FBZ center (boundaries). Therefore, combinations of $C_{\rm FZC}$ and $C_{\rm FZB}$ yield a variety of Floquet insulators with integer $C_{\rm Flq}$ ranging from $-2$ to $2$ [Fig.~\ref{fig:phase}], as its static counterpart permits insulators with $C=0,\pm 1$ [Fig.~\ref{fig:static}]. This is a generic feature of the phase diagrams, allowing Floquet insulators with $|C_{\rm Flq}| \leq 2$, when the mass term is driven in a time-periodic fashion, irrespective of the exact drive protocol. Consequently, it is now also conceivable to find trivial Floquet insulators with $C_{\rm Flq}=0$, which support topological edge modes at same momentum near the center and boundaries of the FBZ.

In Floquet insulators, nondissipative edge modes cross zero and/or $\pm \omega/2$ quasienergies at the Floquet-Bloch band inversion momentum (${\bf K}^{\rm Flq}_{\rm inv}$) when its occurs at the FBZ center and/or boundaries, respectively. It can be anchored from the band structure of the time evolution operator $U(\mathbf{k}, T)$ in a semi-infinite system with $k_x$ or $k_y$ as a good quantum number in terms of its eigenmodes $| \mu_n \rangle$ with quasienergy $\mu_n$, together satisfying $U(\mathbf{k}, T) | \mu_n \rangle =\exp(-i \mu_n T) | \mu_n \rangle$. Explicit results are shown in Figs.~S3-S5 of the SI. Notice that the global phase diagrams [Fig.~\ref{fig:phase}] support insulators with identical $C_{\rm Flq}$ and total number of edge modes (set by the winding number ${\mathcal W}$~\cite{Flq:Thr3}) that respond distinctly to translational symmetry breaking in the bulk of a Floquet crystal (sensitive to ${\bf K}^{\rm Flq}_{\rm inv}$) realized by introducing dislocations or grain boundaries in the system. Therefore, together the Floquet Chern number, winding number and responses to grain boundaries, a direct probe of ${\bf K}^{\rm Flq}_{\rm inv}$, provide a complete classification of Floquet insulators. Gapless modes at the grain boundaries appear whenever ${\bf K}^{\rm Flq}_{\rm inv} \cdot {\bf b}=\pi$ (modulo $2 \pi$) is satisfied.

As an illustrating example, consider two sets of Floquet insulators realized for (a) $(m_1,\omega)=(-9,10)$, $(m_2,\omega)=(6,7)$ and $(m_0,\omega)=(-0.5,5)$, and (b) $(m_1,\omega)=(-11,10)$, $(m_2,\omega)=(3,8)$ and $(m_0,\omega)=(3,5)$, respectively for the kick, step and sinusoidal drives. See Fig.~\ref{fig:phase}. Notice that $C_{\rm Flq}=1$ and ${\mathcal W}=1$ for both sets, and as far as the known topological invariants are concerned ($C_{\rm Flq}$ and ${\mathcal W}$) there is no distinction between them. However, only the set (a) features grain boundary modes, while the set (b) is devoid of it. This is so because the Floquet Bloch band inversion takes place near the $\Gamma$ point of the FBZ for (b), thus yielding ${\bf K}^{\rm Flq}_{\rm inv} \cdot {\bf b}=0$. By contrast, for (a) ${\bf K}^{\rm Flq}_{\rm inv}$ is at the ${\rm M}$ point near the FBZ boundary, yielding ${\bf K}^{\rm Flq}_{\rm inv} \cdot {\bf b}=\pi$. Hence, these two sets of dynamic TIs are distinct phases of matter, which can only be resolved by grain boundary defects. Similar conclusions hold for $C_{\rm Flq}=-1,\pm 2$ Floquet TIs, which can be seen from the phase diagrams in Fig.~\ref{fig:phase}. In these phase diagrams, we also find Floquet insulators that by virtue of featuring the Floquet Bloch band inversion at the ${\rm M}$ point near the center and boundary of the FBZ simultaneously, thus yielding $C_{\rm Flq}=0$, support both normal and anomalous extended grain boundary modes (marked by $\oplus$). See, for example, Floquet insulators at $(m_1,\omega)=(-2,2)$, $(m_2,\omega)=(-0.5,2)$ and $(m_0,\omega)=(-3.5,2)$ for the kick, step and sinusoidal drives, respectively.

\begin{figure*}[t!]
	\includegraphics[width=1.00\textwidth]{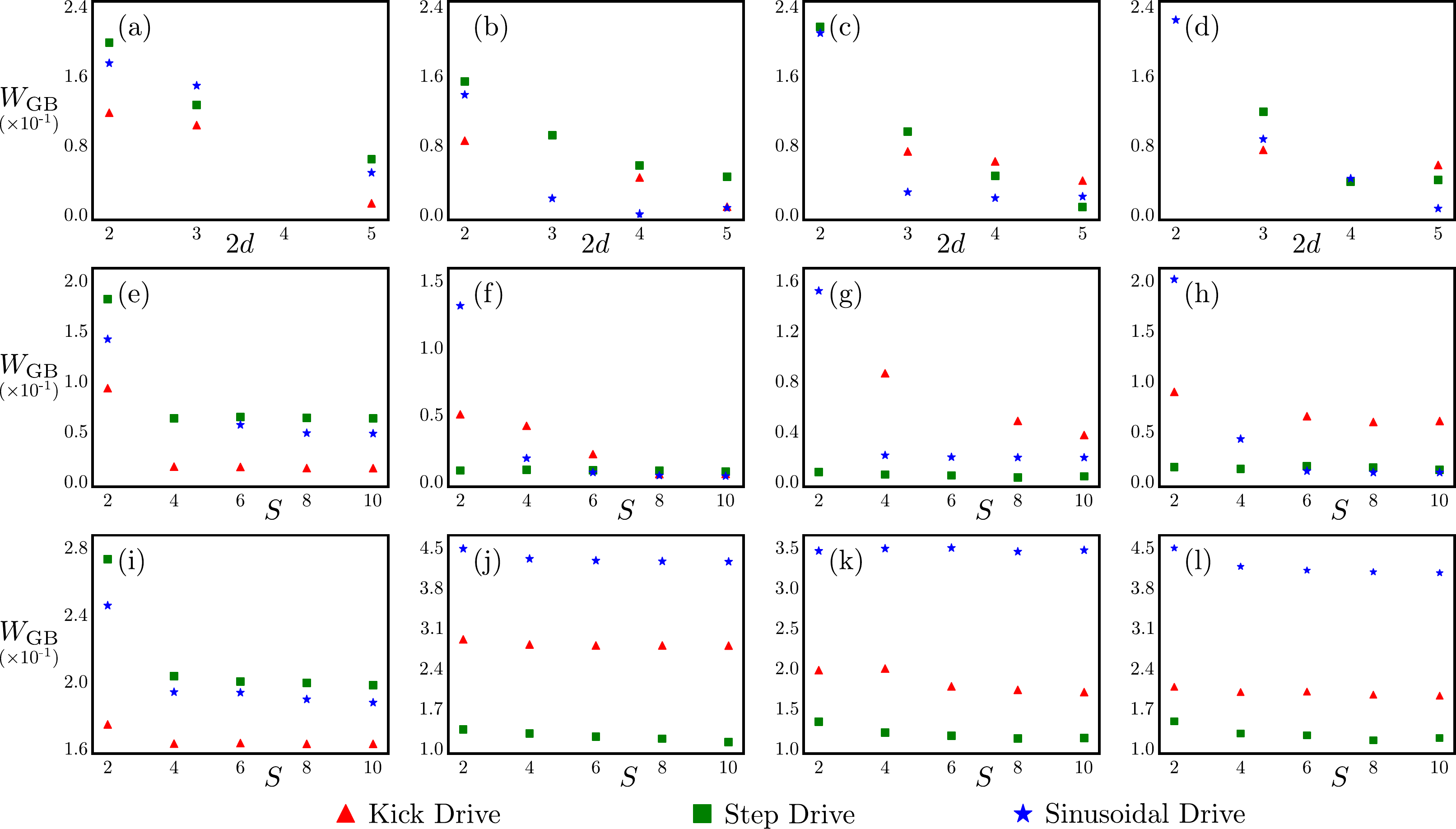}
	\caption{Variation of the band width ($W_{\rm GB}$) of the pure normal [(a), (e) and (i)], pure anomalous [(b), (f) and (j)], and the normal [(c), (g) and (k)] and the anomalous [(d), (h) and (l)] components of the mixed grain boundary dispersive modes with (a)-(d) the separation between two successive (anti)dislocation cores in (anti)grain boundary $2d$ (in units of $a$) for a fixed distance between the ends of the grain and anti-grain boundary defects $S=10 a$, and with $S$ for small angle with $d=5a$ [(e)-(h)] and large angle with $d=2a$ [(i)-(l)] grain boundary defects for kick, step and sinusoidal drives, respectively [Eq.~\eqref{eq:kickdrive}-\eqref{eq:sinusoidal}]. Therefore,	$W_{\rm GB}$ increases with decreasing $2d$ and $S$. At a few missing data points, we encounter numerical instability in Python. The parameter values are the same as in Fig.~\ref{fig:kick}, Fig.~\ref{fig:step} and Fig.~\ref{fig:sinusoidal} for the kick, step and sinusoidal drives, respectively.  
	}~\label{fig:BandwidthFloquetmetal}
\end{figure*}

The requisite topological criteria for the existence of 1D dynamic dispersive bands along the grain boundary in Floquet crystals extends directly from its counterpart in the static systems, but in terms of ${\bf K}^{\rm Flq}_{\rm inv}$. For example, when the Floquet-Bloch band inversion occurs at the ${\rm M}$ point and Floquet zone center (boundary), normal (anomalous) dynamic 1D dispersive fermionic modes appear along such line defects, pinned around the zero ($\pm \omega/2$) quasienergy as the unitary time evolution operator satisfies $\Theta^{-1}U({\bf k},T) \Theta=U({\bf k},T)$. On the other hand, when the Floquet-Bloch band inversion occurs at the ${\rm M}$ point and simultaneously near the center and boundaries of the FBZ, grain boundaries harbor mixed 1D dynamic dispersive states, an admixture of both normal and anomalous gapless modes. These outcomes are insensitive to the drive protocol. We arrive at qualitatively similar outcomes for the kick [Fig.~\ref{fig:kick}], step [Fig.~\ref{fig:step}] and sinusoidal [Fig.~\ref{fig:sinusoidal}] drives. Although, here we present these results for SAGBs, qualitatively similar conclusions hold for large angle grain boundaries as well (see SI). In all these cases, the static system corresponds to a NI, such that the emergence of 1D dynamic dispersive grain boundary modes can solely be attributed to the periodic drives.

Note that inclusion of a constant particle-hole symmetry breaking term $\tau_0 \tilde{m}_0$ in the static Hamiltonian $H_{\rm stat}$ causes an overall shift of all the energy eigenvalues, without affecting its eigenvectors and the resulting topology, as $[\tau_0, {\boldsymbol \tau}]=0$. Similarly, in Floquet systems such a term causes an overall shift of all the quasienergies along the frequency axis by $\omega_0=\tilde{m}_0$, without affecting the eigenmodes $| \mu_n \rangle$ of the time evolution operator, as $[\exp(-i \tilde{m}_0 t), U(\mathbf{k}, t)]=0$ for any $t$ and $V(t)$ [see Eq.~\eqref{eq:unitarydefinition}], yielding the eigenvalue equation at the stroboscopic time  
\begin{eqnarray}
U_0(T) U(\mathbf{k}, T) | \mu_n \rangle = U_0(T) \exp(-i \mu_n T) | \mu_n \rangle \nonumber \\
\equiv \exp(-i T \left[\tilde{m}_0 +\mu_n \right]) | \mu_n \rangle,
\end{eqnarray}
where $U_0(T)=\exp(-i T \tilde{m}_0)$. All our conclusions remains unaffected but within a shifted FBZ of width $\omega$ within frequency range $(-\omega/2+\omega_0,\omega/2+\omega_0)$. However, such a constant shift is unimportant as the FBZ is always defined within the range $(-\omega/2,\omega/2)$ up to an overall translation in the frequency axis due to the time translational symmetry. In the SI, we show the stability of the grain boudnary modes against more general particle-hole symmetry breaking term of the form $\tau_0 [\tilde{m}_0 + \tilde{t}_0 \{ \cos(k_x a)+\cos(k_y a) \}]$, for which $[U_0(\vec{k},T), U(\mathbf{k}, T)]=0$, where $U_0(\vec{k},T)=\exp[-i T \tau_0 d_0(\vec{k})]$. Results are shown in Table~S1 and Fig.~S8 of the SI.

The robustness of the topological criterion, namely ${\bf K}^{\rm Flq}_{\rm inv} \cdot {\bf b}=\pi$ (modulo $2\pi$), for the emergence of the grain boundary modes, is further substantiated by considering a trivial static insulator, subject to time periodic hopping amplitude between the orbitals of opposite parities realized by taking $t_1 \to t_1 (t)$ in the $\mathbf{d}(\vec{k})$-vector from Eq.~\eqref{eq:dvector}. Resulting Floquet phases have no analog in the static system, as topology is insensitive to the magnitude of $t_1$ therein. We consider time periodic kick, step and sinusoidal variations of $t_1(t)$, and show that the phase diagrams support Floquet insulator with a wide range of the Floquet Chern numbers ($C_{\rm Flq}$). Even in such system, dynamic dispersive grain boundary modes emerges at the FBZ center and/or boundaries whenever the condition ${\bf K}^{\rm Flq}_{\rm inv} \cdot {\bf b}=\pi$ (modulo $2\pi$) is satisfied near it. The results are shown in Fig.~S7 of the SI.

The band width of the dispersive grain boundary states ($W_{\rm GB}$) is tunable by its characteristic angle $\theta$ or equivalently $2d$, the distance between two successive (anti)dislocation cores within the extended (anti)grain boundary defect. A smaller $d$ implies a larger $\theta$, and a stronger hybridization between the individual localized (anti)dislocation modes. Therefore, as $d$ ($\theta$) is reduced (increased), $W_{\rm GB}$ increases in Floquet crystals. See Fig.~\ref{fig:BandwidthFloquetmetal}(a)-(d). The same conclusion holds in the static system as shown in Fig.~S1(e) of the SI. We also note that as the distance between the ends of the grain and anti-grain boundaries ($S$) is reduced, the hybridization between the dispersive modes bound to individual extended line defect gets stronger. As a result $W_{\rm GB}$ increases as $S$ is reduced and becomes comparable to a few lattice spacing. These outcomes for small and large angle grain boundaries are shown in panels (e)-(h) and (i)-(l) of Fig.~\ref{fig:BandwidthFloquetmetal}, respectively.

\emph{Discussions}.~To summarize, here we show that irrespective of the drive protocol, Floquet insulators accommodate topologically robust one-dimensional gapless dispersive fermionic modes along grain boundaries whenever inversion of the underlying Floquet-Bloch bands takes place at a finite momentum (${\bf K}^{\rm Flq}_{\rm inv}$) that together with the Burgers vector of the constituting dislocations (${\bf b}$) satisfies ${\bf K}^{\rm Flq}_{\rm inv} \cdot {\bf b}=\pi$ (modulo $2 \pi$). Otherwise, such dissipationless dispersive states appear near zero and/or $\pm \omega/2$ quasienergies when in our model ${\bf K}^{\rm Flq}_{\rm inv}=(\pi,\pi)/a$ near the Floquet zone center and/or boundary. These conclusions hold for both small angle [Figs.~\ref{fig:kick}-~\ref{fig:sinusoidal}] as well as large angle grain boundaries (see SI). Furthermore, as our findings rests on a robust topological criterion (the ${\bf K}^{\rm Flq}_{\rm inv} \cdot {\bf b}$ rule), it should be applicable to driven systems belonging to arbitrary symmetry class in arbitrary dimensions, including Floquet topological superconductors hosting dispersive Majorana modes at the core of grain boundary, dislocation loops resulting from combinations of edge and screw dislocations to name a few, which are left for future investigations. For example, the model in Eq.~\eqref{eq:dvector} also describes a lattice-regularized $p+ip$ paired state for which the two-component Nambu spinor reads $\Psi^\top(\vec{k})=[c(\vec{k}),c^\dagger(-\vec{k})]$ with $t_1$ as the pairing amplitude~\cite{Qi-Zhang-RMP, reedgreen}. In such a system, the ${\boldsymbol \tau}$ matrices operate on the Nambu or particle-hole index, and any particle-hole anisotropy term proportional to $\tau_0$ is forbidden due to the conserved fundamental charge conjugation symmetry. All our results for the dynamically generated dispersive grain boundary modes therefore directly apply to such a system, where these modes are constituted by Majorana fermions.

In Floquet crystals grain boundary modes are (a) dispersive, shown from their Fourier transformation, (b) extended, showing LDOS along the entire defect, and (c) gapless (see the quasi-energy spectra), altogether suggesting their metallic nature. These states are robust against weak disorder as shown in Fig.~S6 of the SI. Therefore, in the future it will be worthwhile computing the transport quantities associated with the dispersive fermionic grain boundary modes from existing formalism~\cite{FloquetTransport1:Th, FloquetTransport2:Th, FloquetTransport3:Th, FloquetTransport4:Th}, which can also be measured in experiments~\cite{FloquetTransport1:Exp}. Robustness and responses of the grain boundary modes, when the Flqouet crystal is coupled to external dissipative baths is yet another avenue to explore in the future~\cite{FloquetBath1:Th}.

With the recent progress in engineering Floquet topological phases in driven quantum crystals~\cite{FlqExpQM}, cold atomic lattices~\cite{FlqExpCA:1, FlqExpCA:2} and dynamic classical metamaterials~\cite{FlqExpMM:1, FlqExpMM:2, FlqExpMM:3, FlqExpMM:4, FlqExpMM:5, FlqExpMM:6} our proposal should be within the reach of currently achievable experimental setups, where dislocation lattice defects~\cite{hamasi:dislocation, nayak:dislocation, jian:dislocation} as well as grain boundaries~\cite{grainboundaryExp:1} have already been demonstrated as tools to detect topological phases of matter in terms of robust modes bound to them. In quantum materials the grain boundary modes are fermionic in nature, whereas metamaterials habor their classical analogues. While defects are ubiquitous in quantum crystals, they can be engineered externally by suitably arranging optical waveguides~\cite{defectsMM:1} and mechanical resonators~\cite{defectsMM:2} in photonic and mechanical lattices, respectively. Each dislocation defect, and subsequently an array of it (grain boundaries), can be created by removing a line of microwave resonators or photonic waveguides or electrical nodes up to the dislocation center, and the requisite $\pi$ hopping phase across the line of missing sites can be manipulated locally~\cite{defectsMM:2, defectsMM:3, defectsMM:4}. The hallmark LDOS of the topological grain boundary modes can be detected via scanning tunneling microscope in quantum crystals, two point pump probe or reflection spectroscopy in photonic lattices and mechanical susceptibility in acoustic lattices, and their gapless nature can be established from fast Fourier transformations, as shown in this work. As the number of (anti)dislocation cores in (anti)grain boundary defects is much smaller than the system size they are not expected to produce any appreciable additional heating effects. Given that Floquet topological phases have already been observed on a number of platforms despite potential heating effects~\cite{FlqExpQM, FlqExpCA:1, FlqExpCA:2, FlqExpMM:1, FlqExpMM:2, FlqExpMM:3, FlqExpMM:4, FlqExpMM:5, FlqExpMM:6}, the predicted extended defect modes should also be observed in these systems.

\noindent 
{\bf Acknowledgments}\\
This work was supported by NSF CAREER Grant No.\ DMR-2238679 of B.R. We are thankful to Vladimir Juri\v ci\' c for discussions.

\noindent 
{\bf Data availability}\\
The datasets used and/or analysed during the current study available from the corresponding authors and Daniel J. Salib (djs421@lehigh.edu) on reasonable request. Main codes and the data for generating the figures presented in the main text and Supplementary Information are already available at https://doi.org/10.5281/zenodo.10819547.

\noindent 
{\bf Author contributions}\\
B.R.\ conceived, structured, and supervised the project and wrote the manuscript. D.J.S.\ performed all the numerical calculations.

\noindent 
{\bf Competing interests}\\
The authors declare no competing interests.

\end{document}